# Fourier-transform infrared spectroscopy with undetected photons from high-gain spontaneous parametric down-conversion

Kazuki Hashimoto[1,*], Dmitri B. Horoshko[2], Mikhail I. Kolobov[2], Yoad Michael[3], Ziv Gefen[3], and Maria V. Chekhova[1,4]

[1] Max Planck Institute for the Science of Light, Staudtstr. 2, 91058 Erlangen, Germany

[2] Univ. Lille, CNRS, UMR 8523 - PhLAM - Physique des Lasers Atomes et Molécules, F-59000 Lille, France

[3] Raicol Crystals, Hamelacha 22, Rosh Ha'Ayin 4809162, Israel

[4] Friedrich-Alexander Universität Erlangen-Nürnberg, Staudtstr. 7/B2, 91058 Erlangen, Germany

* kazuki.hashimoto@mpl.mpg.de

## Abstract

Fourier-transform infrared spectroscopy (FTIR) is an indispensable analytical method that allows label-free identification of substances via fundamental molecular vibrations. However, the sensitivity of FTIR is often limited by the low efficiency of mid-infrared (MIR) photodetectors. SU(1,1) interferometry has previously enabled FTIR with undetected MIR photons via spontaneous parametric down-conversion in the low-parametric-gain regime, where the number of photons per mode is much less than one and sensitive photodetectors are needed. In this work, we develop a high-parametric-gain SU(1,1) interferometer for MIR-range FTIR with undetected photons. Using our new method, we demonstrate three major advantages: a high photon number at the interferometer output, a considerably lower photon number at the sample, and improved interference contrast. In addition, we analyze different methods to broaden the spectral range of the interferometer by aperiodic poling and temperature gradient in the gain medium. Exploiting the broadband SU(1,1) interferometer, we measure and evaluate the MIR absorption spectra of polymers in the 3-μm region.

## Introduction

Fourier-transform infrared spectroscopy (FTIR) is used in various applications such as environmental gas sensing[1] or medical diagnosis[2] as an analytical tool that allows label-free measurements of samples via fundamental molecular vibrations. It obtains broadband mid-infrared (MIR) spectra via time-domain linear interferograms measured with a Michelson interferometer illuminated by a thermal source (e.g., Globar)[3]. Improving the measurement sensitivity of FTIR is important for accurately identifying and analyzing samples, but it is hampered by the low sensitivity and high noise of MIR photodetectors compared with visible/near-IR photodetectors. To address this issue, some research groups have developed upconversion MIR spectroscopy[4–6] with nonlinear frequency conversion or field-resolved spectroscopy with electro-optic sampling[7,8]. However, these methods often require high-power MIR pulsed lasers, dispersion compensation schemes, and/or other high-power lasers for frequency conversion, which makes the spectroscopy system complex.

SU(1,1) interferometry[9] with a narrowband pump source has been recently applied to FTIR as a simple approach to measuring MIR spectra without directly detecting MIR photons. In this technique, nondegenerate spontaneous parametric down-conversion (SPDC) in a nonlinear crystal is used to generate correlated pairs of signal (e.g., visible) and idler (e.g., MIR) photons. A second

SPDC process then occurs sequentially, forming a nonlinear interferometer. The absorption of idler photons between the two crystals affects the signal interference at the interferometer output, which enables measuring the idler absorption without detecting the idler photons ('spectroscopy with undetected photons')[10–13]. FTIR with undetected photons has been implemented with SPDC in the low-parametric-gain regime[11,13], where the number of photons per mode is much less than one and relatively sensitive detectors are needed at the output. Alternatively, if the SU(1,1) interferometer operates under a high parametric gain[14–17], exploiting an intense pulsed laser as a pump source, photon pairs generated by the first crystal seed SPDC in the second one. This new version of spectroscopy with 'undetected photons' has several major advantages. First, it results in higher output powers, sufficient even for a photodetector with moderate sensitivity. Second, due to the amplification in the second crystal, the MIR beam on the sample can be weak enough to interrogate the sample non-invasively. Finally, the interference visibility has a nonlinear dependence on idler amplitude transmittance and can be high even under strong absorption. After demonstrating Fourier-transform near-infrared (NIR) spectroscopy with undetected photons using a high-parametric-gain SU(1,1) interferometer[18], we are now extending the wavelength region to MIR and evaluating the system in this region, which is a significant step for the FTIR spectrometer.

Here, we demonstrate MIR-range FTIR with undetected photons exploiting a high-parametric-gain SU(1,1) interferometer. First, we characterize the bright twin beams (signal: ~0.6 μm, idler: ~3 μm) generated via high-parametric-gain SPDC in a periodically poled potassium titanyl phosphate (ppKTP) crystal. In addition, we measure and evaluate the signal interferogram generated from the interferometer to show the advantages of the high-parametric-gain SU(1,1) interferometer. To increase the sensing bandwidth, we evaluate two spectral broadening methods: a temperature gradient in the ppKTP crystal and an aperiodically poled KTP (apKTP) crystal. Using the developed FTIR spectrometer, we measure broadband MIR spectra of polymer films within a spectral bandwidth of ~300 cm$^{-1}$ and assign some MIR absorption peaks that appear in the spectra.

## Results
### Experimental setup

We develop a high-parametric-gain SU(1,1) interferometer to demonstrate FTIR with undetected photons (Figure 1). A 532-nm 15-ps pulsed laser with a repetition rate of 1 kHz (PL2210A-1K-SH/TH, Ekspla) is focused into a $\chi^{(2)}$ nonlinear crystal using an $f = 200$ mm lens. The pump polarization is adjusted with a half-wave plate (HWP). We use ppKTP or apKTP (Raicol Crystals) as the nonlinear crystal. Both crystals are 10-mm long and designed for type-0 phase matching. The poling period of the ppKTP crystal is fixed at 13.3 μm, and that of apKTP gradually increases from 12.3 to 14.0 μm along the crystal. The crystal temperature is adjusted with an oven and/or a Peltier element. The high-parametric-gain SPDC inside a nonlinear crystal generates bright twin beams (signal: visible, idler: MIR). The signal and pump pulses, reflected by a dichroic mirror (DM), travel through the reference arm and return to the same path after being reflected by an $f = 100$ mm spherical mirror. The idler pulse goes to the scan arm and is collimated with an $f = 100$ mm CaF$_2$ lens. The collimated idler pulse passes through the test sample, is reflected by an end mirror, and travels back to the same path. The end mirror is placed on a motorized stage for scanning the optical path-length difference (OPD) between the reference and scan arms. The three pulses are again focused into the nonlinear crystal for a second SPDC process. Inside the crystal, the signal pulse is amplified or de-amplified depending on the pump phase relative to the sum of signal and idler phases introduced by the motorized stage. The resultant signal interferogram contains information on the idler absorption because the absorption changes the number of amplified signal photons. After the crystal, the signal pulse is separated from the pump pulse with a DM and collimated with an $f = 100$ mm lens. The residual pump after the DM is removed by long-pass filters. The collimated signal pulse is detected with a Si power meter (S130VC, Thorlabs) with 15-Hz radio-frequency (RF) low-pass filtering. The recorded signal interferogram is sampled with an 8-bit digitizer (USB-5133, National Instruments) with a sampling rate of 1.53 kSamples/s. The interferogram is RF bandpass filtered to remove the DC part and the unwanted high- and low-frequency noises. The OPD of the AC-coupled interferogram (FTIR interferogram) is calibrated with a simultaneously measured continuous-wave (CW) interferogram with a helium-neon (HeNe) laser. Fourier-transforming the calibrated FTIR interferogram yields the FTIR spectrum.

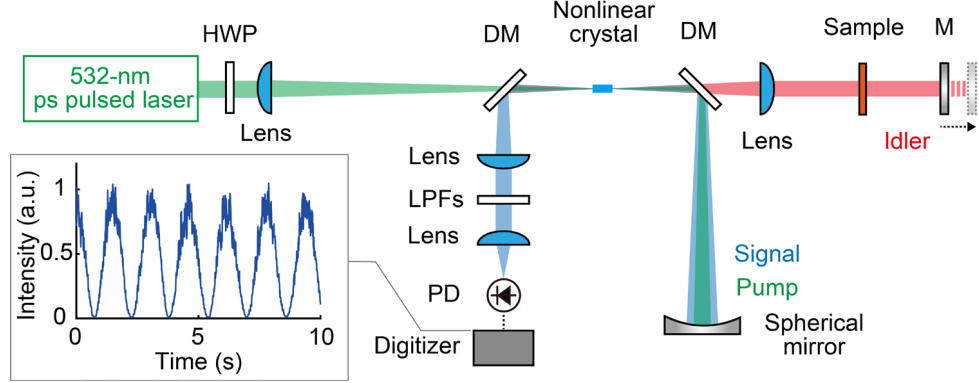

**Figure 1:** Schematic of the SU(1,1) interferometer for FTIR with undetected photons. HWP: half-wave plate, DM: dichroic mirror, M: mirror, LPFs: long-pass filters, PD: photodetector. Either ppKTP or apKTP is used as the nonlinear crystal for the interferometer. The inset shows a signal interferogram generated by the ppKTP crystal without any sample, which is measured with the Si power meter and recorded with the software for the console (PM100D, Thorlabs).

## FTIR interferogram from an SU(1,1) interferometer

The FTIR interferogram obtained with the SU(1,1) interferometer represents the delay-dependent part of the total number of signal photons detected at the interferometer output as a function of the idler delay time $\tau$ and, for a quasi-monochromatic pump at circular frequency $\omega_p$, is given by[18]

$$\Delta N(\tau) = \int_{\omega_p/2}^{\omega_p} I(\omega_s) T_i(\omega_p - \omega_s) \cos[(\omega_p - \omega_s)\tau + \rho(\omega_s)] \frac{d\omega_s}{2\pi}, \quad (1)$$

where $\omega_s$, $T_i(\omega_p - \omega_s)$, and $\rho(\omega_s)$ denote the signal angular frequency, the intensity transmittance of the sample at the idler frequency $\omega_i = \omega_p - \omega_s$, and the phase acquired due to a dispersive propagation in the crystal and sample, respectively. $I(\omega_s)$ is proportional to the spectral density of signal photon flux at frequency $\omega_s$ after the second passage through the crystal. In our experiments, its value is obtained by taking the FTIR interferogram without the sample. Integration in Equation (1) is over the entire signal band. A Fourier transform of Equation (1) gives the complex FTIR spectrum

$$F(\omega_i) = \frac{1}{2} I(\omega_p - \omega_i) T_i(\omega_i) e^{-i\rho(\omega_p - \omega_i)}, \quad (2)$$

where $\omega_i$ is limited to the idler band $[0, \omega_p/2]$. The transmittance spectrum of the sample is obtained as $T_i(\omega_i) = |F(\omega_i)|/|F_0(\omega_i)|$, where $F_0(\omega_i)$ is measured without the sample.

## ppKTP crystal with temperature tuning

First, we evaluate the signal spectrum generated from the ppKTP crystal via the first SPDC process. Figure 2(a) shows the signal spectra measured with a visible spectrometer (AvaSpec-ULS3648-USB2, Avantes) with a spectral resolution of 1.4 nm at several crystal temperatures from 25 to 195 °C. The central wavelength of the signal spectrum changes from 637 nm (15700 cm$^{-1}$) to 626 nm (15970 cm$^{-1}$) with temperature tuning. The intensity drop at shorter wavelengths is due to the idler absorption originating from the second overtone bands of the fundamental vibrations of KTP[19]. The idler absorption inside the crystal reduces the number of signal photons generated by high-parametric-gain SPDC. We also measure the average signal power dependence on the pump power at a crystal temperature of 22 °C (Figure 2(b)). The number of photons per mode generated by the SPDC process is theoretically described as $N = \sinh^2 r$, where $r$ denotes the parametric gain being proportional to the square root of the pump power. The signal power exponentially increases with the pump power and saturates above ~400 µW. The signal power obtained at a pump power of 400 µW is around 8 µW. The unsaturated data are fitted with the function $y = A \sinh^2(B\sqrt{x})$, where $A$=5.1 x 10$^{-19}$ and $B$=25 are the fitting coefficients. Considering the fitting curve, we achieve $r$~16 at a pump power of 400 µW. Note that the signal spatiotemporal structure should be single mode when strictly measuring the parametric gain, but here, we check it without the mode selection.

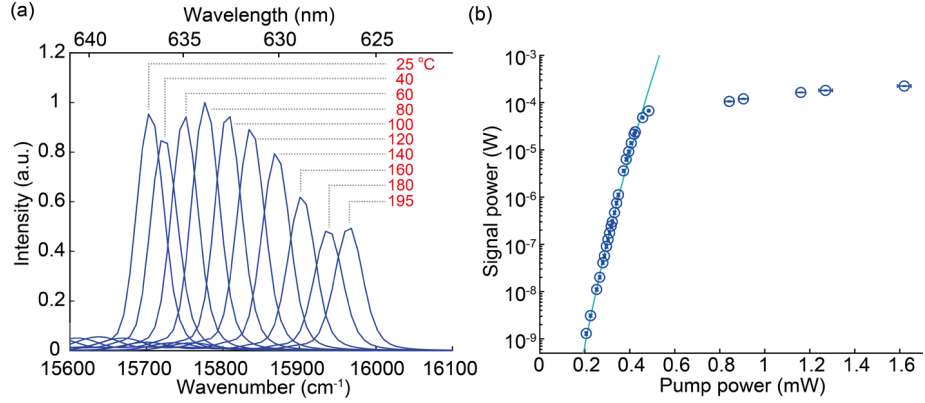

**Figure 2:** (a) Signal spectra generated by the first SPDC process in the ppKTP crystal for several crystal temperatures. The signal spectrum is tunable over 270-cm$^{-1}$-width by controlling the temperature. (b) Signal power vs. pump power. The blue points are the measurements, and the sky-blue line is the curve fitting the first 10 data points. The fitting function is $y = A \sinh^2(B\sqrt{x})$, with $A$=5.1 x 10$^{-19}$ and $B$=25 being the fitting parameters. The X- and Y-error bars indicate the standard deviation of the pump and signal powers, respectively.

Next, we measure the FTIR interferogram and spectrum using the SU(1,1) interferometer with the ppKTP crystal. Figure 3(a) shows the FTIR interferogram measured at a crystal temperature of 21 °C. As shown in the inset, we observe a sinusoidal waveform at the center of the interferogram. Figure 3(b) shows the FTIR spectra obtained by Fourier-transforming the 3-times averaged interferograms at temperatures from 21 to 190 °C. The full-width at half-maximum (FWHM) width of the FTIR spectrum at 21 °C is 14 cm$^{-1}$. The Fourier modulus $|F(\omega_i)|$ exponentially increases as the pump power increases and saturates at a pump power of ~250 µW (See Supplementary Information). At a pump power of 210 µW, the mean power of the signal detected with the Si power meter is 580 nW (580 pJ), while the idler power at the sample is expected to be around only 0.2 nW (0.2 pJ). The wavenumber (wavelength) varies from 3109 cm$^{-1}$ (3216 nm) to 2855 cm$^{-1}$ (3503 nm) by tuning the temperature.

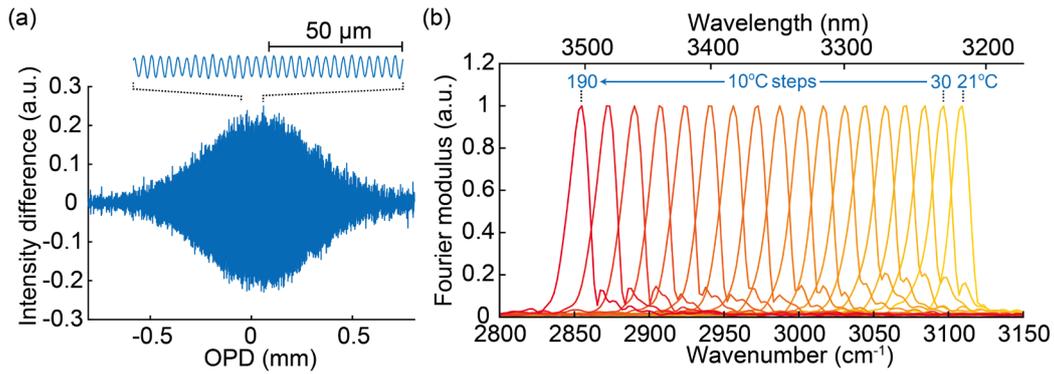

**Figure 3:** (a) A single-shot FTIR interferogram obtained by the SU(1,1) interferometer with the ppKTP crystal. The inset shows the zoomed view of the interferogram. (b) The FTIR spectra at crystal temperatures of 21 - 190°C. Each spectrum is normalized to the peak Fourier modulus.

Using the frequency tuning scheme, we measure the MIR transmittance spectra of polymer films. Figure 4 shows the (intensity) transmittance spectra of a polystyrene film and a plastic wrap measured with the SU(1,1) interferometer (red points) and the reference spectra measured with a standard FTIR spectrometer (Varian 670-IR, Varian) (dotted gray lines) with a spectral resolution of 16 cm$^{-1}$. The Fourier modulus measured with the SU(1,1) interferometer is proportional to the double-pass idler amplitude transmittance, which is equal to the intensity transmittance, entering Equation (2). To obtain the transmittance, we integrate the Fourier moduli of the FTIR spectra (with and without the samples) over a bandwidth of 17 cm$^{-1}$. The MIR transmittance spectra measured with our method agree well with the reference spectra. We observe fundamental molecular vibrations of polystyrene in the measured IR spectrum, for example, CH$_2$ asymmetric stretching at 2923 cm$^{-1}$ and aromatic CH stretchings at 3016 cm$^{-1}$ and 3057 cm$^{-1}$ [20]. Also, we observe large absorption bands in the plastic-wrap spectrum and attribute them to C-H stretching modes.

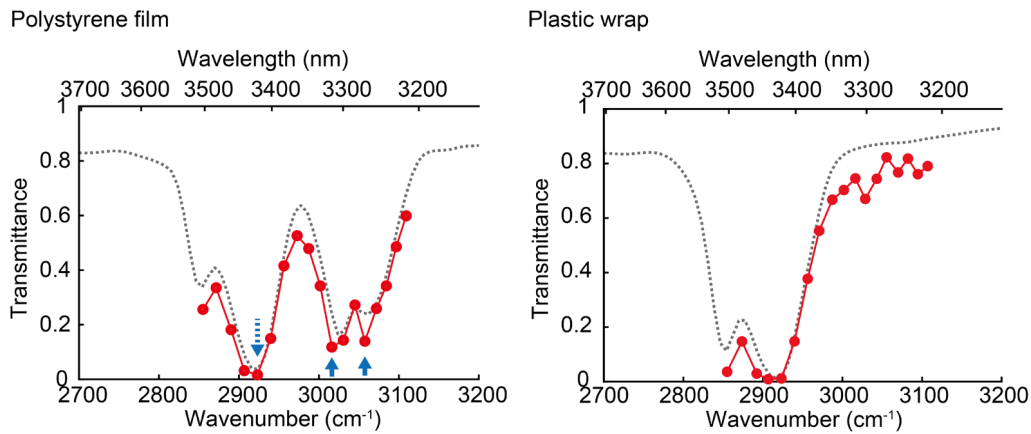

**Figure 4:** MIR (intensity) transmittance spectra of a polystyrene film (left) and a plastic wrap (right). The red points are measured with the SU(1,1) interferometer and the dotted gray lines, with a standard FTIR spectrometer with a spectral resolution of 16 cm$^{-1}$. The solid and dotted arrows show the aromatic CH stretching and CH$_2$ asymmetric stretching modes, respectively.

We further characterize the visibility of the signal interferograms. The interferograms are measured with the power meter and recorded with the software for the power meter console. Figure 5(a) shows the signal interferogram measured at (double-pass) idler and signal amplitude transmittances of $|t_i|$ =0.39 and $|t_s|$ =0.80, respectively, with a visibility of 67%. The visibility is evaluated as $V = (J_{max} - J_{max})/(J_{max} + J_{max})$, where $J_{max}$ and $J_{min}$ denote the maximum and minimum intensities of the interferograms after eliminating the high-frequency noises by smoothing. Meanwhile, the same transmittances would lead to much lower visibility if the interference were measured at a low parametric gain. Figure 5(b) shows the simulated interferograms[15] at $|t_i| = 0.39$ and $|t_s| = 0.80$ with a parametric gain of 11 (high-parametric-gain regime) and 0.1 (low-parametric-gain regime). The high-gain simulated interferogram has visibility close to the experimental one (79%), but the low-gain visibility is much lower (38%). This is because, while the visibility in the low-parametric-gain regime is linear in $|t_i|$, $V_L = \frac{2|t_s|}{1+|t_s|^2}|t_i|$, the visibility in the high-parametric-gain regime is nonlinear in $|t_i|$, $V_H = \frac{2|t_s|}{|t_i|^2+|t_s|^2}|t_i|$.

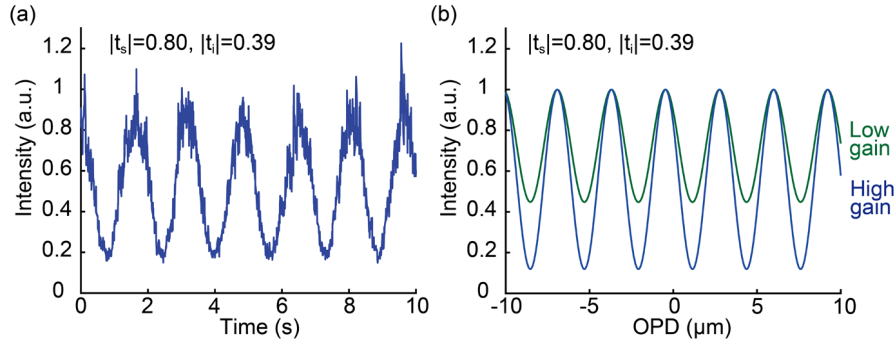

**Figure 5:** (a) A signal interferogram measured at $|t_i| = 0.39$ and $|t_s| = 0.80$. The visibility is 67%. (b) Simulated signal interferograms at $|t_i| = 0.39$ and $|t_s| = 0.80$ with a parametric gain of 11 (blue) and 0.1 (green). The visibilities at the high- and low-parametric-gain regimes are 79% and 38%, respectively. Each interferogram is normalized to the peak intensity.

**ppKTP crystal with a temperature gradient**

One approach to spectral broadening is to apply a temperature gradient[21] inside the nonlinear crystal, which gradually changes the phase-matching wavelength along the crystal. To create the temperature gradient inside the ppKTP crystal, we place one side of the crystal on a plate heated by an oven while the other side is cooled by a Peltier cooler. In this configuration, roughly 5 mm, 2 mm, and 3 mm out of 10 mm are cooled, uncontrolled, and heated, respectively. The temperature of the oven is set at 140°C and the temperature of Peltier, at 15°C. Figure 6 shows the signal spectrum generated via the first SPDC process inside the ppKTP crystal with a temperature gradient at a pump power of 344 μW. The FWHM bandwidth of the spectrum is 170 cm⁻¹ (from 15700 to 15870 cm⁻¹). The peaks at 15730 and 15830 cm⁻¹ are generated from the cool and hot parts of the ppKTP crystal, respectively.

Further, with the SU(1,1) interferometer spectrum broadened by the temperature gradient, we measure the FTIR spectrum. Figure 7(a) shows a 10-times averaged FTIR interferogram measured without any sample. The signal power on the photodetector is around 4 μW at a pump power of 438 μW. We obtain the FTIR spectrum by Fourier-transforming the FTIR interferogram (Figure 7(b)). The FWHM spectral bandwidth is 113 cm⁻¹ (from 2971 cm⁻¹ to 3084 cm⁻¹), eight times larger than for the ppKTP crystal with uniform temperature. The bandwidth at -10-dB intensity level is 151 cm⁻¹.

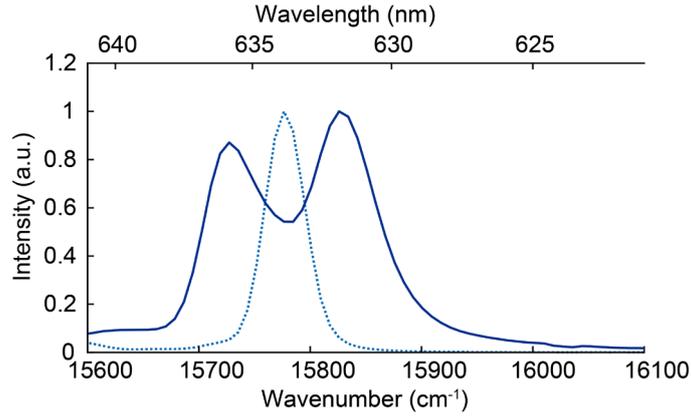

**Figure 6:** Signal spectra generated via the first SPDC process in the ppKTP crystal with a temperature gradient (solid blue) and a fixed temperature of 80°C for comparison (dotted blue).

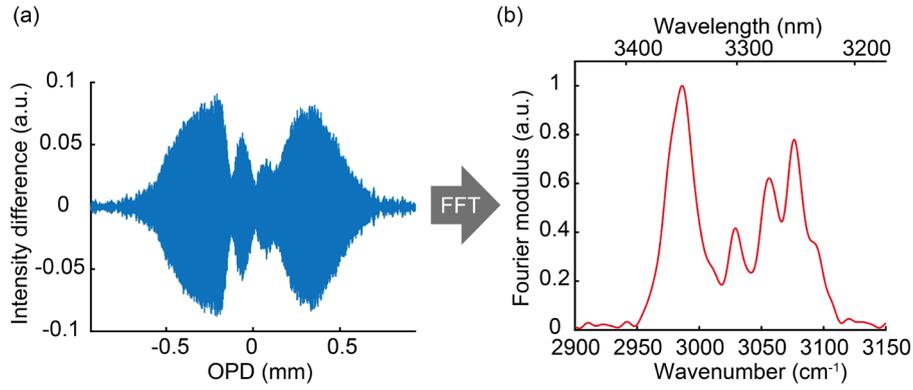

**Figure 7:** (a) A 10-times averaged FTIR interferogram measured with the temperature gradient scheme. FFT: fast Fourier transform. (b) The FTIR spectrum obtained by Fourier transforming the interferogram shown in (a).

## apKTP crystal

We explore an additional method for increasing the spectral bandwidth of the twin beams. We design and fabricate a KTP crystal whose poling period is chirped along the crystal (apKTP crystal). The grating vector of the apKTP crystal (defined by $2\pi/\Lambda$, where $\Lambda$ denotes the poling period) is linearly chirped from 510 rad/mm ($\Lambda = 12.3$ μm) to 449 rad/mm ($\Lambda = 14.0$ μm) with a chirp rate of $\zeta = 6.1$ rad/mm$^2$. The chirped grating vector is designed such that the idler spectrum covers the C-H, N-H, and O-H bands of molecules[22] (from 2700 to 3700 cm$^{-1}$). Figure 8(a) shows the grating vector and the poling period profiles used for the mask design. The linear chirp is realized with a piecewise-constant quasi-linear profile with a local period increment of 25 nm. The microscopic images of the fabricated apKTP crystal are shown in Figure 8(b).

Figure 8(c) shows the twin-beam SPDC spectra from the apKTP crystal, calculated numerically with the equations obtained by some of us earlier[23,24] with the Rosenbluth parameter of $\nu = 1$. The Rosenbluth parameter for the linear chirp is defined as $\nu = |\gamma|^2/\zeta$, where $\gamma$ is the three-wave mixing coupling constant, proportional to the pump amplitude. From the calculation, we estimate the FWHM spectral bandwidth of the SPDC spectra as 870 cm$^{-1}$ (signal: 15150 cm$^{-1}$ - 16020 cm$^{-1}$, idler: 2780 cm$^{-1}$ - 3650 cm$^{-1}$). We note the increment of 25 nm of the piecewise constant structure has negligible effects on the spectral shape.

Next, we measure the broadband signal spectrum generated from the apKTP crystal (under the first SPDC process) with the visible spectrometer. In this measurement, we use $f = 300$ mm lens to focus the pump pulse into the crystal. Figure 9 compares the measured and the simulated signal spectra. The simulated spectrum is obtained by convolving the spectrum calculated with $\nu = 0.6$ with a 1.4-nm-width Gaussian function to match the spectral resolution. The measured spectrum spanning from 15170 cm$^{-1}$ to 16070 cm$^{-1}$ agrees well with the simulated one. At higher gain, we observed the bandwidth-narrowing effects and sidebands at 580-620 nm and around 550 nm, which are discussed in Supplementary Information.

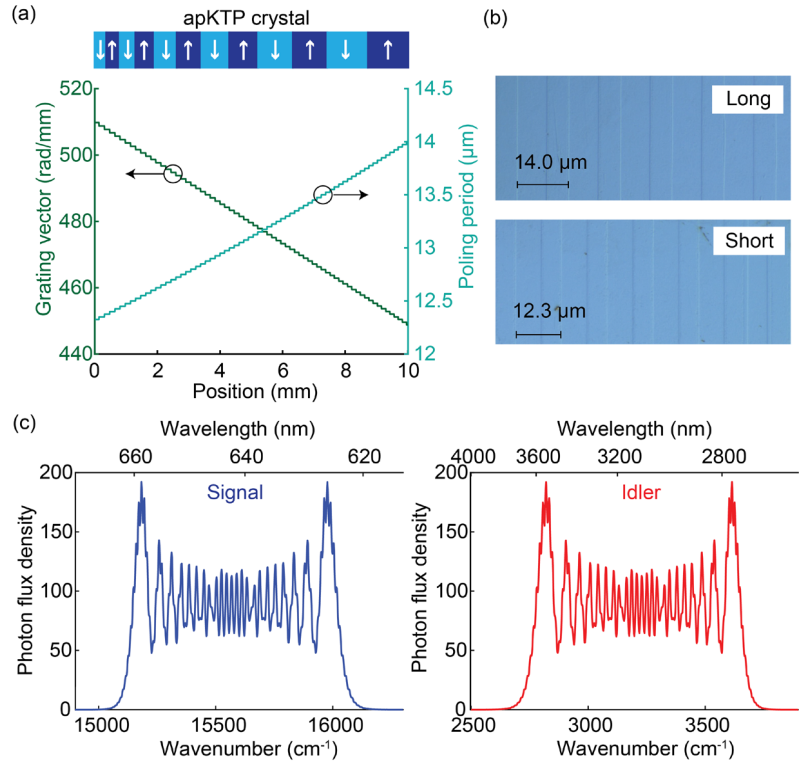

**Figure 8:** (a) Grating vector and poling period of the apKTP crystal. (b) Microscopic images of the long- and short-poling-period regions of the apKTP crystal. (c) Numerically calculated twin-beam spectra generated via the first SPDC process in an apKTP crystal. The Rosenbluth parameter[23] is $\nu = 1$.

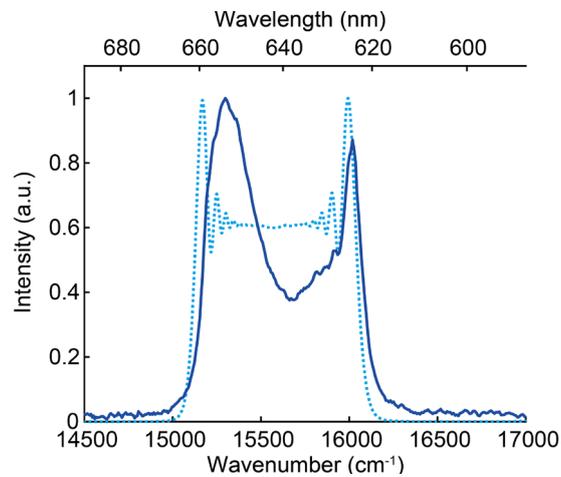

**Figure 9:** Measured (solid blue) and simulated (dotted sky-blue) signal spectra generated via the first SPDC process in the apKTP crystal.

Finally, we implement the apKTP crystal in the SU(1,1) interferometer to demonstrate broadband FTIR with undetected photons. The pump is focused in the crystal by a lens with a focal length of 200 mm, and its power is 1.37 mW. The crystal is kept at 150°C to mitigate the signal-intensity decrease that occurred at room temperature with high pump power. We may attribute the intensity decrease to the gray tracking effects inside the KTP crystal[25,26]. Figure 10(a) shows the 36-times averaged FTIR interferograms with (lower) and without (upper) a polystyrene film. Fourier-transforming the interferograms yields the FTIR spectra with and without the sample, as shown in Figure 10(b). The spectrum taken with the sample contains several absorption peaks originating from the molecular vibrations of polystyrene. The spectral bandwidth of the FTIR spectrum without the sample evaluated at -10-dB Fourier-modulus level is 300 cm$^{-1}$ (the FWHM bandwidth is 170 cm$^{-1}$). We also observed the bandwidth narrowing in the FTIR spectra, probably because of the narrowing effects mentioned in the previous paragraph. Figure 10(c) shows the MIR (intensity) transmittance spectra of a polystyrene film measured with the SU(1,1) interferometer (solid red) and the reference spectra (dotted gray) with a spectral resolution of 5 cm$^{-1}$. We clearly obtain the absorption lines of polystyrene. The transmittance mismatch between our method and the standard FTIR spectrometer could be caused by the surface quality of the sample, degrading the SU(1,1) interference.

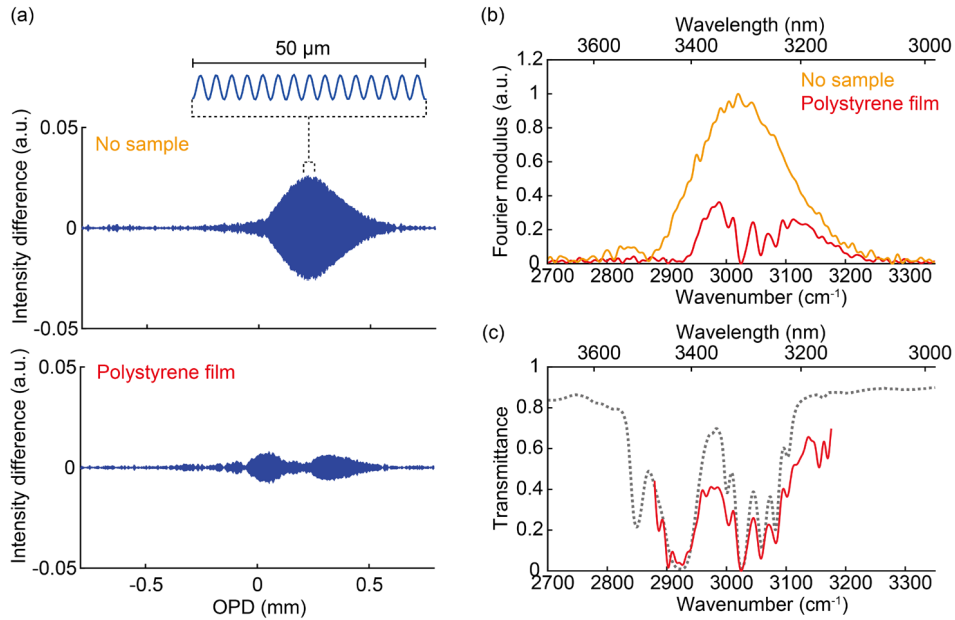

**Figure 10:** (a) FTIR interferograms measured by an SU(1,1) interferometer with an apKTP crystal. The upper and lower panels show the interferogram with and without a polystyrene film. The inset shows the enlarged view of the interferogram without the sample. (b) The FTIR spectra obtained by Fourier-transforming the interferograms shown in (a) (Red: with the sample, orange: without the sample). (c) MIR (intensity) transmittance spectra of a polystyrene film. The solid-red and dotted-gray lines show the transmittance spectra measured with our system and a standard FTIR spectrometer with a spectral resolution of 5 cm$^{-1}$, respectively.

## Discussion

The spectral range of the SPDC spectra generated from the 10-mm ppKTP crystal with the temperature control is 270 cm$^{-1}$, and that of the 10-mm apKTP crystal is more than three times broader (900 cm$^{-1}$). Increasing/decreasing the temperature (e.g., to above 200°C and/or below 10°C) of the ppKTP crystal leads to expanding the spectral range in principle. In the apKTP crystal case, the range can be extended by increasing the chirp rate of the grating vector and/or increasing the crystal length while keeping the chirp rate. The spectral bandwidth of the FTIR spectrum obtained with the apKTP crystal is 300 cm$^{-1}$, 30% of the SPDC spectral bandwidth generated via the first SPDC process. We assume the optimization of the chirp function can improve the bandwidth of the signal spectrum at the interferometer output.

The idler pulse energy in the sample in the ppKTP crystal case is estimated to be ~0.2 pJ (~0.2 nW), while the amplified signal at the interferometer output is 580 pJ (580 nW). The strong parametric amplification allows for interrogating a sample non-invasively with a lower energy idler pulse and detecting the molecular vibration information with a higher energy signal pulse, sufficient for detecting with a cost-effective moderate sensitivity detector. The signal pulse energy in the temperature gradient case is ~4 nJ (~4 μW), also detectable with a normal sensitivity photodetector. In the apKTP crystal case, the signal pulse energy is around 10 pJ (10 nW), corresponding to the photon number per coherence time of ~$10^5$, being orders of magnitude larger than the low-parametric-gain case[13] (roughly $10^{-9}$ photons per coherence time). In addition, the measurement time of a single-shot interferogram in our system with the apKTP crystal is ~40 s at a maximum OPD of 1.6 mm, much shorter than the latter case with the same OPD (several hours). Note that we roughly compare the two methods without normalizing signal-to-noise ratio (SNR) and other measurement conditions.

The maximum visibility of the signal interferogram is 98%, shown in the inset of Figure 1. With an idler amplitude transmittance of 0.39, the visibility is 67%, higher than in the low-parametric-gain case (38%). The high-visibility feature allows for sufficiently utilizing the detector's dynamic range when measuring the samples with large absorption.

We evaluate the SNR of the FTIR spectra measured with the undetected photon schemes. In the ppKTP crystal case, the single-shot SNR of the 14-cm$^{-1}$-width FTIR spectrum obtained at a maximum OPD of 1.6 mm is 230, which is evaluated by the ratio between the peak Fourier modulus and the standard deviation where no peaks exist. The SNR of the 36-times averaged 170-cm$^{-1}$-width FTIR spectrum measured with the apKTP crystal with a maximum OPD of 1.6 mm is 105. The noise linearly increases with the signal power on the photodetector (See. Supplementary Information), i.e., the SNR does not depend on the signal power. Therefore, the SNR of the system is currently dominated by the signal pulses' relative intensity noise (RIN), which can be alleviated by stabilizing the pump intensity and/or increasing the number of spatiotemporal modes.

Our method can advance with further modifications of the developed system. Using other nonlinear crystals (e.g., AgGaS$_2$ (AGS) crystal[27,28]), we can also perform FTIR with undetected photons in the molecular fingerprint region. In addition, our system can acquire hyperspectral MIR images[29] by implementing a camera for detection. The principle of this method is also applicable to MIR optical coherence tomography[30,31], which may be useful for the non-invasive 3D imaging of paintings. Furthermore, our method can be integrated into multimodal spectroscopy[32] techniques that also use a pulsed laser to induce multiple nonlinear optical effects at the same time.

**Conclusion**

In conclusion, we developed a high-parametric-gain SU(1,1) interferometer to demonstrate MIR-range FTIR with undetected photons. To provide a large sensing bandwidth, we used two different strategies. In the first one, we generated SPDC in a periodically poled KTP crystal with a temperature variation, either in sequence or with a gradient along the crystal. In the second one, we used an aperiodically poled (chirped) KTP crystal. The FTIR spectra obtained from the interferometer covered up to

~300 cm$^{-1}$ in the 3-μm region. The measured MIR transmittance spectra of polymer samples agreed well with the reference measurements. We also observed the strong parametric amplification in the second crystal and the high visibility at a large idler loss, which are advantages of our method over the low-parametric-gain version. The amplification enables interrogating a sample non-invasively with weak idler pulses and detecting the absorption information through the amplified signal pulses using a photodetector with moderate sensitivity. The high-visibility interferogram allows us to use the dynamic range of the detector and the digitizer sufficiently. Our system has a great potential to advance the MIR spectroscopy techniques used in various application fields.

## Acknowledgments

We thank Michael Frosz and Azim-Onur Yazici for letting us use the FTIR spectrometer (Varian 670-IR, Varian) and Kyoohyun Kim for letting us use the microscope to inspect the apKTP crystal. K. H. acknowledges the financial support by JSPS (Overseas Research Fellowships). M. V. C., Y. M., and Z. G. acknowledge the support by QuantERA II Programme (project SPARQL), which has received funding from the European Union's Horizon 2020 research and innovation programme under Grant Agreement No 101017733, with the funding organization Deutsche Forschungsgemeinschaft. D. B. H. and M. I. K. acknowledge the support of Agence Nationale de la Recherche (France) via grant ANR-19-QUAN-0001 (QuICHE) and of Franco-Bavarian University Cooperation Center via grant FK-09-2023. Y. M. and Z. G. acknowledge the support by the Israel Innovation Authority.

## Disclosures

The authors declare no competing interests.

## Data availability

The data provided in the manuscript are available from the corresponding author upon reasonable request.

# Supplementary Information: Fourier-transform infrared spectroscopy with undetected photons from high-gain spontaneous parametric down-conversion

Kazuki Hashimoto[1,*], Dmitri B. Horoshko[2], Mikhail I. Kolobov[2], Yoad Michael[3], Ziv Gefen[3], and Maria V. Chekhova[1,4]

[1] Max Planck Institute for the Science of Light, Staudtstr. 2, 91058 Erlangen, Germany

[2] Univ. Lille, CNRS, UMR 8523 - PhLAM - Physique des Lasers Atomes et Molécules, F-59000 Lille, France

[3] Raicol Crystals, Hamelacha 22, Rosh Ha'Ayin 4809162, Israel

[4] Friedrich-Alexander Universität Erlangen-Nürnberg, Staudtstr. 7/B2, 91058 Erlangen, Germany

* kazuki.hashimoto@mpl.mpg.de

**Supplementary Note 1. Fourier modulus vs. pump power**

We measure the Fourier modulus of an FTIR spectrum vs. pump power using the FTIR spectra measured with a ppKTP crystal (Figure S1). The points are obtained from the peak Fourier moduli of the FTIR spectra. The modulus increases exponentially as the pump power increases and saturates. The mean signal power is 580 nW (580 pJ) at a pump power of 210 µW.

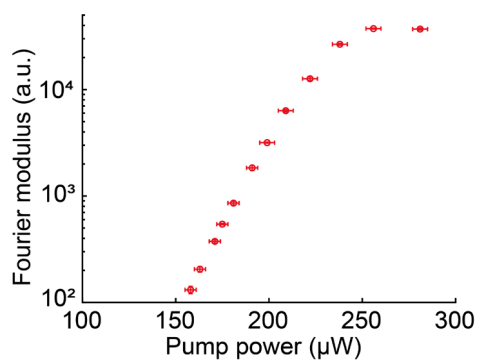

**Figure S1**: The Fourier modulus of an FTIR spectrum vs. pump power evaluated from the FTIR spectra measured with a ppKTP crystal. The X- and Y-axis error bars show the pump power and the modulus standard deviations, respectively.

**Supplementary Note 2. Characterization of SPDC spectra**

The signal spectrum generated from an apKTP crystal contains sidebands at 580-620 nm and around 550 nm other than the target spectral region of 620-660 nm (Figure S2(a)). The upper panel in Figure S2(b) shows the wavevector mismatch curve of the z-polarized wave in the KTP crystal at 22°C calculated from the Sellmeier equations for 0.532-1.1 µm[1], for 1.1-3.4 µm[2,3], and for >3.4 µm[4,5], assuming the derivatives of the refractive index[6] over the temperature from 20 - 25°C are constant. The curve intersects the blue-shaded area (between the highest and the lowest grating vectors) at 620 - 660 nm, 580-590 nm, and ~550 nm. We calculate the signal spectrum using the wavevector mismatch curve (lower panel in Figure S2(b)) with the Rosenbluth parameter[7] of $\nu = 1$ and obtain spectral bands at 620 - 660 nm, 580-590 nm, and ~550 nm. From the calculation, we attribute the sidebands that appeared at 580-620 nm and around 550 nm to the signal spectrum generated by SPDC.

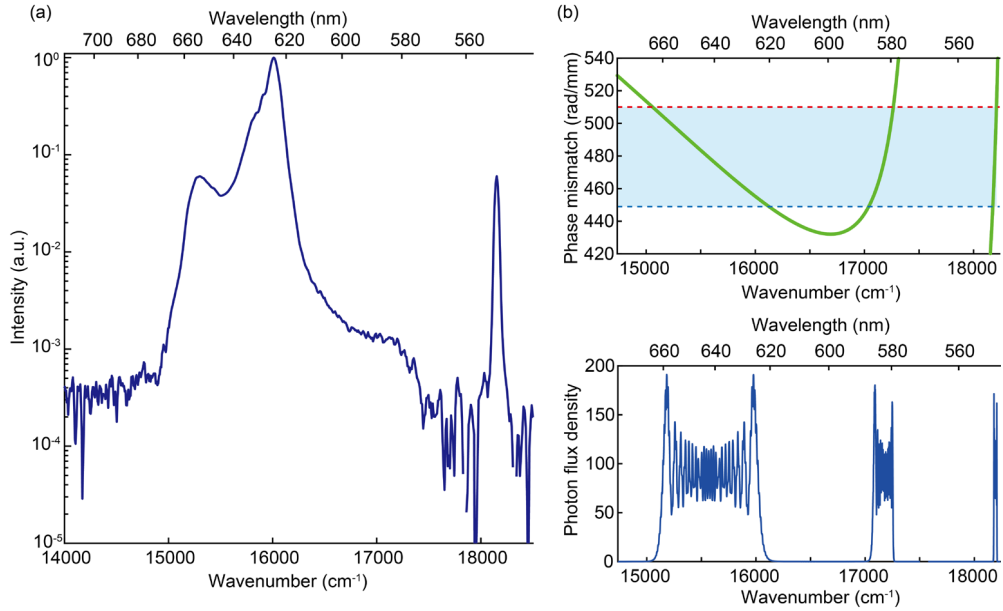

**Figure S2**: (a) Signal spectrum generated by the first SPDC process in an apKTP crystal. (b) Upper: the wavevector mismatch curve of the z-polarized wave in the KTP crystal. Dotted red and blue lines show the highest and lowest grating vectors for the apKTP crystal, respectively. Lower: the simulated signal spectrum with the Rosenbluth parameter of $\nu = 1$.

We observe bandwidth-narrowing effects in the signal spectra measured with an apKTP crystal, especially at a higher pump power. Figure S3 shows the signal spectra at various pump powers in two cases: (1) the pump pulses enter from the crystal side with the shortest poling period (12.3 μm) and exit from the other side with the longest poling period (14.0 μm) (Short-Long), and (2) the other way around (Long-Short). The focal length of the lens for focusing the pump pulses is 300 mm. The spectral intensity increases with the pump power in both cases, but the intensity at the shorter wavelength (~630 nm) increases faster than at the longer one (~650 nm) in the Short-Long case, where the longer wavelength signals are generated near the beginning of the crystal. As shown in Figure S3, the tendency reverses in the Long-Short case. The unbalanced signal intensity causes the bandwidth-narrowing effects. We attribute these effects to the propagation loss of the signal photons generated near the beginning of the crystal during the propagation and the further signal depletion at a higher pump power due to some nonlinear effects inside the crystal. The FTIR spectral bandwidth would also be decreased due to the narrowing.

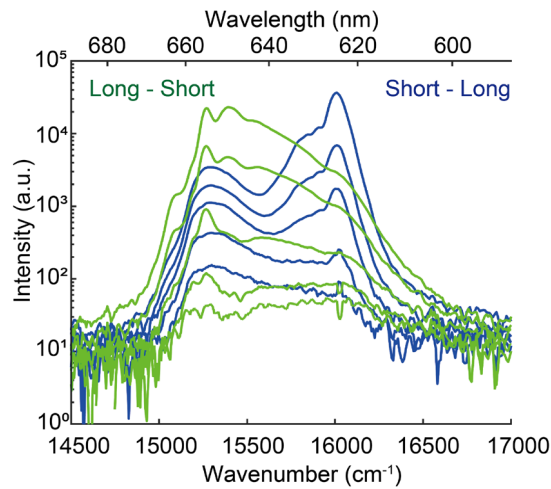

**Figure S3**: Signal spectra with various pump powers at different crystal directions (Blue: Short-Long, Light green: Long-Short). The pump powers for the spectra measured under the Short-Long case are 1.9, 2.4, 2.8, 3.1, and 3.3 mW from the bottom plot, and those under the Long-Short case are 1.5, 1.9, 2.3, 2.6, and 2.8 mW from the bottom plot.

**Supplementary Note 3. SNR and dominant noise**

We evaluate the dominant noise of the system by measuring the noise vs. the mean signal power on the power meter, as shown in the upper panel in Figure S4. The noise, evaluated by the standard deviation of the signal power at the DC (i.e., non-oscillatory) part of the interferogram measured with an apKTP crystal, is linear in the mean power. Therefore, the dominant noise of the measurement is the relative intensity noise (RIN) of the signal pulses. The RIN-limited SNR is unchanged with the mean signal power on the detector. Indeed, the SNR obtained from the ratio of the mean power to the noise remains constant to the mean signal power, as shown in the lower panel.

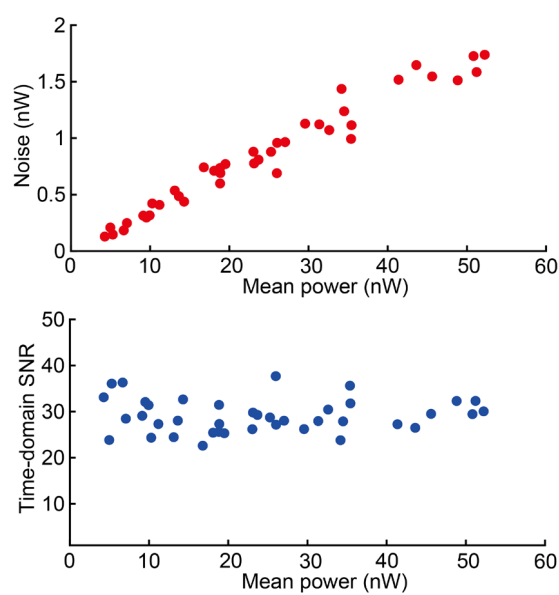

**Figure S4**: Noise evaluation of the signal interferograms. Upper: noise dependence on the mean signal power detected with the power meter at the interferometer output. Lower: SNR vs. mean signal power. The SNR is evaluated from the ratio of the mean power to the noise.